\documentclass{aa} 
\usepackage[varg]{txfonts}
\usepackage{soul}
\usepackage{ifthen}
\usepackage{graphicx}
\usepackage{multirow}
\usepackage{bbm}

\newcommand{\for}[1]{Eq.~(\ref{#1})}

\newcommand{\fig}[1]{Fig.~\ref{#1}}
\newcommand{\figFull}[1]{Figure~\ref{#1}}
\newcommand{\tab}[1]{Table~\ref{#1}}
\newcommand{\sect}[1]{Sect.~\ref{#1}}

\usepackage{hyperref}
\hypersetup{colorlinks=true, citecolor=blue, linkcolor=blue, urlcolor=blue}

\usepackage{xcolor}
\definecolor{darkred}{HTML}{660022}

\newcommand{\upp}[1]{^{(#1)}}

\newcommand{\OmegaM}{\Omega_\mathrm{m}}
\newcommand{\sigEig}{\sigma_8}
\newcommand{\wZero}{w_0^\mathrm{de}}

\newcommand{\gala}{\mathrm{gal}}
\newcommand{\noise}{\mathrm{noise}}
\newcommand{\peak}{\mathrm{peak}}
\newcommand{\pix}{\mathrm{pix}}

\newcommand{\rmi}{\mathrm{i}}   
\newcommand{\rmG}{\mathrm{G}}   

\defcitealias{Liu_etal_2014}{Liu, X. et al.}
\defcitealias{Liu_etal_2015a}{Liu, X. et al.}
\defcitealias{Liu_etal_2016}{Liu, X. et al.}
\defcitealias{Liu_etal_2014a}{Liu, J. et al.}
\defcitealias{Liu_etal_2015}{Liu, J. et al.}
\defcitealias{Lin_Kilbinger_2015}{Paper I}
\defcitealias{Lin_Kilbinger_2015a}{Paper II}
\defcitealias{Lin_etal_2016a}{Paper III}
\newcommand{\PaperI}{\citetalias{Lin_Kilbinger_2015}}
\newcommand{\PaperII}{\citetalias{Lin_Kilbinger_2015a}}
\newcommand{\PaperIII}{\citetalias{Lin_etal_2016a}}
\newcommand{\Camelus}{\textsc{Camelus}}
\newcommand{\LiuFourteenA}{\citetalias{Liu_etal_2014} (\citeyear{Liu_etal_2014})}
\newcommand{\LiuFourteenB}{\citetalias{Liu_etal_2014a} (\citeyear{Liu_etal_2014a})}
\newcommand{\LiuFifteenA}{\citetalias{Liu_etal_2015} (\citeyear{Liu_etal_2015})}
\newcommand{\LiuFifteenB}{\citetalias{Liu_etal_2015a} (\citeyear{Liu_etal_2015a})}
\newcommand{\LiuSixteen}{\citetalias{Liu_etal_2016} (\citeyear{Liu_etal_2016})}

\begin{document}

\title{Quantifying systematics from the shear inversion on weak-lensing peak counts}
\titlerunning{Inversion systematics on weak-lensing peak counts}
\author{Chieh-An Lin\inst{\ref{inst1},\ref{inst2},\ref{inst3}} \and Martin Kilbinger\inst{\ref{inst2},\ref{inst4}}}
\authorrunning{C.-A. Lin \& M. Kilbinger}
\institute{
        Institute for Astronomy, University of Edinburgh, Royal Observatory, Blackford Hill, Edinburgh EH9 3HJ, UK\label{inst1}\\
\email{\texttt{calin@roe.ac.uk}}
        \and 
        Service d'Astrophysique, CEA Saclay, Orme des Merisiers, B\^at 709, 91191 Gif-sur-Yvette, France\label{inst2}
        \and 
        Fenglin Veteran Hospital, 2 Zhongzheng Rd. Sec. 1, Fenglin Township, Hualien 97544, Taiwan\label{inst3}
        \and 
        Sorbonne Universités, UPMC Univ. Paris 6 et CNRS, UMR 7095, Institut d'Astrophysique de Paris, 98 bis bd Arago 75014 Paris, France\label{inst4}
}
\date{Received 27 March 2017 / Accepted 13 February 2018 / Conformed to the published version}

\abstract
        {Weak-lensing peak counts provide a straightforward way to constrain cosmology by linking local maxima of the lensing signal to the mass function. Recent applications to data have already been numerous and fruitful. However, the importance of understanding and dealing with systematics increases as data quality reaches an unprecedented level. One of the sources of systematics is the convergence-shear inversion. This effect, inevitable when carrying out a convergence field from observations, is usually neglected by theoretical peak models. Thus, it could have an impact on cosmological results. In this paper, we study the bias from neglecting (mis-modeling) the inversion. Our tests show a small but non-negligible bias. The cosmological dependence of this bias seems to be related to the parameter $\Sigma_8\equiv(\OmegaM/(1-\alpha))^{1-\alpha}(\sigEig/\alpha)^{\alpha}$, where $\alpha=2/3$. When this bias propagates to the parameter estimation, we discovered that constraint contours involving the dark energy equation of state can differ by 2$\sigma$. Such an effect can be even larger for future high-precision surveys and we argue that the inversion should be properly modeled for theoretical peak models.}
        

\maketitle

\section{Introduction}
\label{sect:intro}

The observation of background light sources results in distorted images due to light deflection. This effect is called gravitational lensing. The lensing signal tracks the matter distribution of the Universe and provides evidence of cosmic history. In the weak-lensing (WL) regime, the information about structure formation is sensitive to both linear and nonlinear scales, providing a crucial tool to constrain cosmology (see for example \citealt{Kilbinger_2015} for a review). With upcoming large surveys such as Euclid \citep{Laureijs_etal_2011} and the Large Synoptic Survey Telescope (LSST, \citealt{LSSTScienceCollaboration_etal_2017}), WL has been considered as an important probe to find out, notably, the nature of dark energy and the gravity law at cosmic scales.

Peak counts are powerful non-Gaussian statistics that allow us to constrain cosmology from WL. Peaks of high signal-to-noise ratio (S/N) are shown to be good tracers of massive halos (\citealt{Yang_etal_2011}; \citealt[][hereafter \PaperI]{Lin_Kilbinger_2015}). The more massive halos there are, the more likely it is to find high S/N peals in a lensing map. While older studies \citep{Kruse_Schneider_1999, Kruse_Schneider_2000, Reblinsky_etal_1999, Bartelmann_etal_2002, Hamana_etal_2004, Wang_etal_2004} seem to only focus on reducing false positive and false negative detections, recent studies (\citealt{Jain_VanWaerbeke_2000, Dietrich_Hartlap_2010, Kratochvil_etal_2010, Fan_etal_2010}; \PaperI\ among others) tend to model true and false detections together in order to probe cosmology. Indeed, WL peak counts are sensitive to changes of shape of the mass function. In the literature, it has been shown that peak counts alone constrain cosmology more strictly than two-point-correlation functions and the combination of both statistics improves the constraints (\citealt{Dietrich_Hartlap_2010}; \citetalias{Liu_etal_2015} \citeyear{Liu_etal_2015}). Applications to observational data are also multiple. The data of the Canada-France-Hawaii-Telescope Lensing Survey (CFHTLenS) have been analyzed by \LiuFifteenA\ and \LiuSixteen, the CFHT Stripe-82 Survey by \LiuFifteenB, Dark Energy Survey Science Verification (DES-SV) data by \citet{Kacprzak_etal_2016}, and the Kilo-Degree Survey (KiDS) by \cite{Martinet_etal_2018} and \cite{Shan_etal_2018}.

For future large and deep lensing surveys, the importance of dealing with systematics increases significantly. To cite a few, potential sources of systematics are shape measurement errors \citep{Cardone_etal_2014}, photometric redshift uncertainty \citep{Cunha_etal_2014, Bonnett_2016, Choi_etal_2016, Gruen_Brimioulle_2017}, intrinsic alignment of galaxies \citep{Chisari_etal_2014, Codis_etal_2015, Schaefer_Merkel_2015, Schrabback_etal_2015, Krause_etal_2016}, baryon physics \citep{Mohammed_etal_2014, Harnois-Deraps_etal_2015}, and instrumental responses \citep{Gurvich_Mandelbaum_2016, Okura_etal_2016, Kannawadi_etal_2016, Plazas_etal_2016}. Concerning WL peak counts, studies of systematics have been done by \citet{Yang_etal_2013} and \citet{Osato_etal_2015} for baryon physics, \LiuFourteenA\ for masking, and \LiuFourteenB\ for magnification bias (see also \citealt{Lin_2016} for a review of WL-peak-related studies). These existing studies are not sufficient for modeling the peak statistics at high precision.

One of the systematics that has not been addressed is the convergence-shear inversion. In WL, the (reduced) shear is observable but the convergence field is not. In order to obtain the convergence, one common way is to invert its relation to the shear via the lensing potential. This is called the Kaiser-Squires inversion \citep{Kaiser_Squires_1993}. It assumes that the galaxy shape noise could be modeled as a Gaussian random field, which conserves the same properties before and after the inversion. However, in practice, this hypothesis is rarely fulfilled for two reasons. First, the field is probed only on irregular galaxy samples. Second, masks introduce missing data and border effects. Because of these facts, the noise level is not spatially uniform. It can even be far from uniform for the galaxy density and the filter size used in realistic survey scenarios. In addition, inversion methods does not account for the reduced shear without nonlinear effects. As a result, the convergence-shear inversion can lead to a systematic bias to WL observables. Alternatively, \citet{Seitz_Schneider_1995} proposed a nonlinear method to properly take the reduced shear into account. However, this approach still relies on the assumption of uniform noise, which needs to be dealt with carefully.

Up to now, there exist three modeling approaches for predicting WL peak counts. The first approach consists in generating $N$-body runs to simulate structures that cause lensing (for example, \citealt{Kratochvil_etal_2010} and its series including \citetalias{Liu_etal_2015} \citeyear{Liu_etal_2015}). Then, peaks are counted from the derived lensing maps. The second approach is a pure analytical calculation, which computes the probability of different lensing signal levels given a line of sight (for example, \citealt{Fan_etal_2010} and its series including \citetalias{Liu_etal_2015a} \citeyear{Liu_etal_2015a}). After that, the probability of identifying a peak can be deduced with random field theory. The third approach adopts a semi-analytical model (this work). Based on some simple assumptions, it generates simulations in a much faster way than $N$-body runs and count peaks from the resulting lensing maps.

In the literature, \citet{Fan_etal_2010} and \citet{Shirasaki_2017} have proposed theoretical models for WL peaks. These models directly predict peak counts on a convergence field from theory.\footnote{The model of \citet{Fan_etal_2010} can estimate peaks directly from a shear field, but it is computationally expensive.} Therefore, such approaches do not account for the inversion that is required for establishing a convergence field from data. The purpose of this paper is to quantify this modeling bias from theoretical WL peak models when neglecting the inversion. Two algorithms, the Kaiser-Squire and Seitz-Schneider inversions, are considered in this study. We use our stochastic model developed in \PaperI, \citet[][\PaperII]{Lin_Kilbinger_2015a}, and \citet[][\PaperIII]{Lin_etal_2016a} to examine this effect. The stochasticity of the model makes including different inversions and different map-making methods straightforward. This makes our model an ideal tool to achieve our objective.

After the introduction, the theoretical formalism of different inversion methods is introduced in \sect{sect:formalism}. Then, we explain the methodology, including the construction of different comparison cases and the choice of some detailed settings in \sect{sect:methodology}. The results are presented in \sect{sect:results}. Then, we will summarize the paper and conclude with a discussion in \sect{sect:summary}.

\section{Inversion formalism}
\label{sect:formalism}

In this paper, we aim to test the impact of the convergence-shear inversion on WL peak counts. The most commonly used technique is the Kaiser-Squires (KS) inversion \citep{Kaiser_Squires_1993} in which the convergence $\kappa$ is given by
\begin{align}\label{for:KS_inversion}
        \hat{\kappa} = \frac{\ell_1^2 - \ell_2^2 - 2\rmi\ell_1\ell_2}{\ell_1^2 + \ell_2^2} \hat{\gamma}\ \ \ \text{for $\ell_1\ell_2>0$,}
\end{align}
where $\gamma$ is the shear, \ $\hat{\mbox{}}$\ \ the Fourier transform operator, and $\ell$ the Fourier mode. The arguments of $\kappa$ and $\gamma$ are omitted. To recover $\kappa$, the inverse transform leaves an undetermined constant term in direct space, corresponding to $\ell_1=\ell_2=0$. This global constant is usually set to zero as the expected mean of the convergence field is null. In the WL regime, the reduced shear $g\equiv\gamma/(1-\kappa),$ which has an unbiased estimator using the observed galaxy ellipticities \citep{Seitz_Schneider_1997}, is often approximated by the shear, so that $g\approx\gamma$ and \for{for:KS_inversion} is applied on $g$ directly. 

In order to account for the factor $1/(1-\kappa)$ properly, \citet{Seitz_Schneider_1995} proposed an alternative inversion method (SS inversion). It follows the iterative process below:
\begin{align}
        \kappa\upp{0} &= 0, \notag\\
        \gamma\upp{i} &= (1-\kappa\upp{i-1})\frac{1-\text{sign}\Big(\det\mathcal{A}\upp{i-1}\Big)\sqrt{1-|\delta|^2}}{\delta^*} \ \ \ \text{for $i\geq1$,\ \ and} \notag\\
        \hat{\kappa}\upp{i} &= \frac{\ell_1^2 - \ell_2^2 - 2\rmi\ell_1\ell_2}{\ell_1^2 + \ell_2^2} \hat{\gamma}\upp{i} \ \ \ \text{for $i\geq1$}, \label{for:iterative_inversion}
\end{align}
where the index $\upp{i}$ stands for the $i$-th iterate, $\det\mathcal{A} = (1-\kappa)^2 - |\gamma|^2$ and $\delta = 2\gamma(1-\kappa)/((1-\kappa)^2+|\gamma|^2) = 2g/(1+|g|^2)$. In the case where $\det\mathcal{A}$ is always positive (i.e., $|g| < 1$), the second line of \for{for:iterative_inversion} is equivalent to $\gamma\upp{i} = (1-\kappa\upp{i-1})g$. The KS and SS inversions will be the two methods to study in this paper.

\section{Methodology}
\label{sect:methodology}

We used the \Camelus\ code (proposed in \PaperI) to model peak-count predictions. This semi-analytical model adopted a halo approach, deriving peak number counts from a mass function. It first drew halos with mass and redshift with respect to the input mass function. Then, it randomized the halos' angular positions. After that it assigned source galaxies and computed their respective lensing signal. Finally, the model created a lensing map and counted peaks by their S/N. In this way, massive dark-matter halos were considered as the major source of WL peaks.

Concerning how \Camelus\ carried out halo sampling and galaxy assignment in this study, readers are invited to read \PaperIII\ for details. In the following, we explain how lensing maps were made and how peaks were defined.
In order to test the impact from the convergence-shear inversion, we compared peak counts in three cases:
\begin{itemize}
        \item Case 1: $\kappa$ is computed directly from ray-tracing simulations.
        \item Case 2: $g$ is computed from ray-tracing simulations and $\kappa$ is given by the KS inversion.
        \item Case 3: $g$ is computed from ray-tracing simulations and $\kappa$ is given by the SS inversion.
\end{itemize}
The $\kappa$ and $g$ could be given by a halo's projected mass. Here, $\kappa$ was computed following Eqs. (26) and (27) of \citet{Takada_Jain_2003a} and $g$ by combining these previous equations with Eqs. (16) and (17) of \citet{Takada_Jain_2003b}. The convergence fields of all three cases were generated from the same series of \Camelus\ simulations, so that there was no statistical fluctuation caused by different halo or galaxy samples. In this set-up, Case 1 closely follows the logic of theoretical models from \citet{Fan_etal_2010} and \citet{Shirasaki_2017} where the inversion is omitted, whereas Cases 2 and 3 represented the real-world scenario. If the inversion caused a misestimation of peak counts, then a comparison of Case 1 to Case 2 or 3 should be able to separate this contribution from other systematic sources.

For this paper, we performed two sets of fast simulations. The first was to quantify the bias from mis-modeling the inversion using the following diagnostic: $(N_\peak^\kappa - N_\peak^g) / N_\peak^g$, where $N_\peak^{\kappa,g}$ was the number of peaks obtained directly from the convergence or from the shear with inversion. The denominator was the peaks from the inverted shear since this corresponded to the observation. Therefore, this indicator showed the peak-count deviation due to mis-modeling. The choice of the denominator was due to the fact that the bias that this paper studies does not refer to ``the bias caused by the inversion'' but ``the bias of mis-modeling which does not account for the inversion''. Hereafter, we will use ``mis-modeling bias'' as a shortened term to refer to the bias from mis-modeling of the inversion. For this test, 2000 independent realizations were carried out for nine cosmologies varying $\OmegaM$ and $\sigEig$. The dark energy equation of state was fixed at $\wZero=-0.96$. A realization was a 450$\times$450 map with a pixel width of 0.8 arcmin, such that the field area was 36 deg$^2$. Three Gaussian filters of widths 1.2, 2.4, and 4.8 arcmin were applied. A characteristic mask taken from the W1 field of CFHTLenS data was also applied.

The second set of fast simulations was to measure how cosmological constraints could be affected. Here, we chose to study cosmological constraints in a three-dimensional space composed of $\OmegaM$, $\sigEig$, and $\wZero$. We mimicked the observational data vector by peaks from Case 2 under a reference cosmology $(\OmegaM, \sigEig, \wZero) = (0.28, 0.82, -0.96)$, while the likelihood was computed in both Cases 1 and 2. By doing so, the difference between likelihoods could visualize how mis-modeling would propagate to parameter constraints. This second set was the same fast simulations that were used in \PaperIII, which consisted of 37,536 cosmologies with 500 independent realizations each. The pixel size, field area, and smoothing filters were configured in the same way as the first simulation set.

Weak-lensing peaks from Cases 2 and 3 were defined in the same way as in \PaperIII: shape noise had been added for all galaxies, which were binned later; after the inversion and the filtering, the noise level for S/N was estimated locally based on the effective number of neighboring galaxies covered under the filter, in the same way as Eq. (19) of \PaperIII. However, for Case 1, the simulations have been done differently. For Case 1, we first binned galaxies to generate a noise-free convergence map, then added a constant pixel noise before filtering the map. The noise variance $\sigma_\pix^2$ is defined as
\begin{align}
        \sigma_\pix^2 = \frac{\sigma_\epsilon^2}{2} \frac{1}{n_\gala A_\pix},
\end{align}
where $\sigma_\epsilon^2$ was the sum of the variance of the intrinsic ellipticity distribution, $n_\gala$ the source galaxy number density, and $A_\pix$ the area of a pixel. The motivation for this setting, though different from other cases, is that theoretical models are restricted to such a configuration. Only if this configuration is respected do we start to follow the same modeling logic as them. As a result, the definition of S/N should also change. Since the noise was constant for Case 1, the expected noise level in S/N was simply defined as 
\begin{align}
        \sigma_\noise^2 = \frac{\sigma_\epsilon^2}{2} \frac{1}{2\pi n_\gala \theta_\rmG},
\end{align}
where $\theta_\rmG$ was the smoothing size.

From our cosmological parameter space, we projected constraint contours on all of the two-dimensional plans to study numerically the constraint diagnostic. To take into account degeneracies, the constraint contours were fitted with three indicators,
\begin{align}
        \Sigma_8 &= \left(\frac{\OmegaM+\beta}{1-\alpha}\right)^{1-\alpha} \left(\frac{\sigma_8}{\alpha}\right)^\alpha, \label{for:indicator_1}\\
        I_1 &= \OmegaM - a_1 \wZero,\ \ \ \text{and} \label{for:indicator_2}\\
        I_2 &= \sigEig + a_2 \wZero, \label{for:indicator_3}
\end{align}
which were respectively degeneracy lines for the $\OmegaM$-$\sigEig$, $\sigEig$-$\wZero$, and $\OmegaM$-$\wZero$ planes. This definition of $\Sigma_8$ allowed a good measurement of contour width independent from $\alpha$. Concerning $\wZero$, Eqs. \eqref{for:indicator_2} and \eqref{for:indicator_3} assumed that $\wZero$ was connected to two other considered parameters by an affine relation. An analysis of these indicators, already used in \PaperIII, would allow us to examine how degeneracy lines vary. Here, we set $\alpha=2/3$, $\beta=0$, $a_1=0.108$, and $a_2=0.128$ and computed the fitted $\Sigma_8$, $I_1$, and $I_2$ respectively in Cases 1 and 2. This choice of $\alpha$ and $\beta$ made $\Sigma_8$ a functional form similar to $\sigEig\OmegaM^{0.5}$; and the choices of $a_1$ and $a_2$ were the best-fit results taken from \PaperIII.

\section{Results}
\label{sect:results}

\begin{figure}[tb]
        \centering
        \includegraphics[width=\columnwidth]{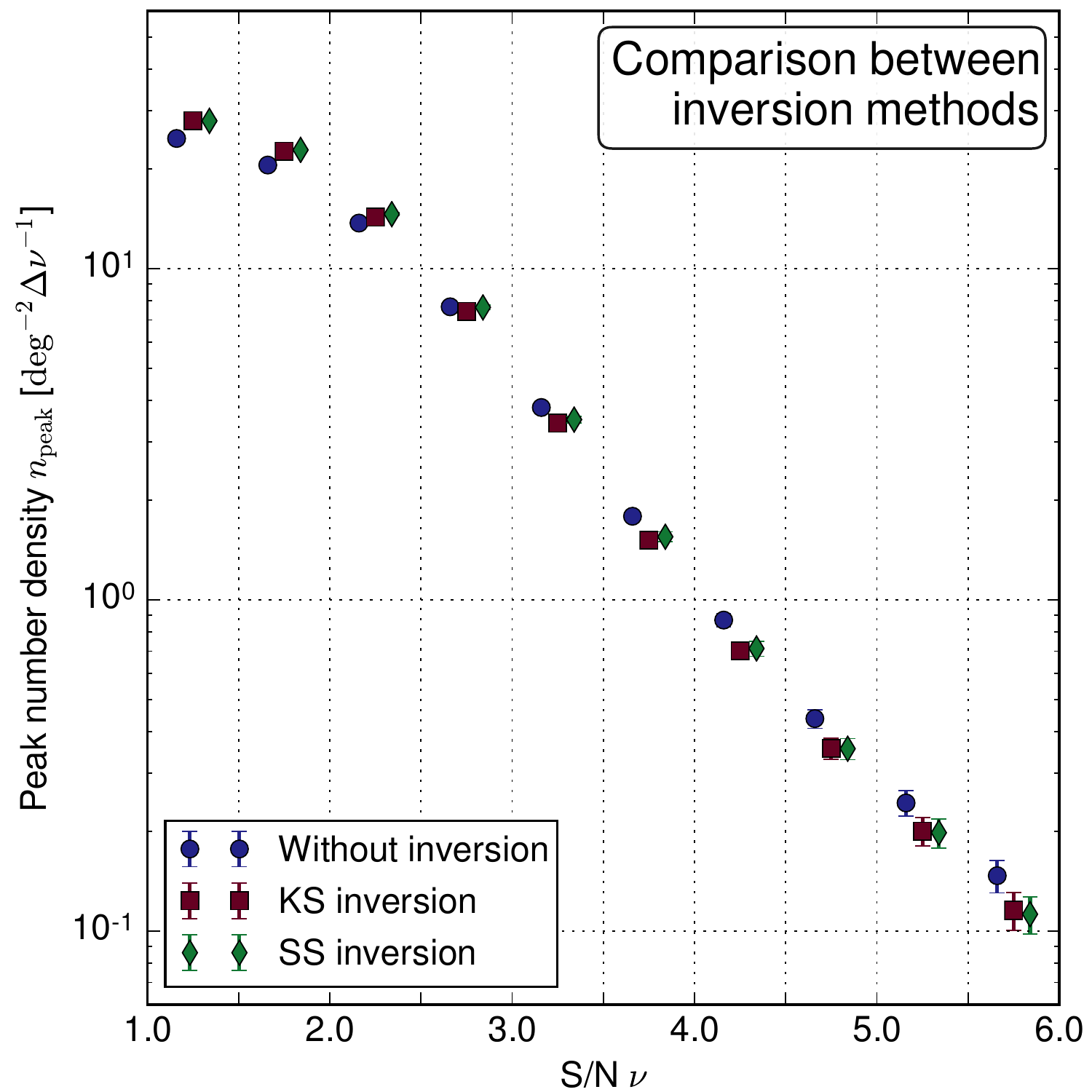}
        \caption{Peak function from different studied cases. Blue circles, red squares, and green diamonds respectively show the mean peak density over 2000 realizations in Cases 1, 2, and 3. The error bars are rescaled to correspond to a survey of 1080 deg$^2$. The filter size is 2.4 arcmin. The input cosmology is $(\OmegaM, \sigEig, \wZero)=(0.28, 0.82, -0.96)$. We can see that the impact of not modeling the inversion is small. Between two inversion techniques, there is only a slight difference.}
        \label{fig:peakHist_compInv}
\end{figure}

\figFull{fig:peakHist_compInv} shows the peak function from the reference cosmology $(\OmegaM, \sigEig, \wZero) = (0.28, 0.82, -0.96)$ with the filter of 2.4 arcmin, using the first simulation set mentioned in \sect{sect:methodology}. The error bars are rescaled to correspond to a survey of 1080 deg$^2$. We see that the biases of the convergence case compared to both shear cases are relatively small. They are globally contained within $\pm$30\% of Cases 2 and 3. The difference between the KS and SS inversions is even smaller, revealing that the inversion method has little importance for  peaks. This figure also shows that the inversion decreases the peak counts under this configuration. However, we find that this cannot be generalized for all cosmologies and all filter sizes.

Interpreting the mechanism of this bias is challenging. One may think that the bias is caused by using $g=\gamma/(1-\kappa)$ in the calculation instead of the true shear $\gamma$. If we follow this reasoning, for high peaks, the reduced shear $g$ should be larger than $\gamma$ since $\kappa$ is positive. After a linear KS inversion, the deduced convergence should also be larger than the true convergence, so the peak function from Case 2 should be situated at the right-hand side (which also means on top) of the one from Case 1. However, this is exactly the opposite of what \fig{fig:peakHist_compInv} indicates, meaning that the reduced shear is not the origin of this bias. Actually, if we compare Case 3 (where the factor $1/(1-\kappa$) has been taken into account) to Case 1, we see that the difference between two peak functions is still present. As a result, we suggest the irregularity of galaxy distribution to be the major origin of this bias.

\begin{table}[tb]
        \centering
        \caption{Inversion mis-modeling bias obtained under different cosmologies and different inversion methods. The upper panel is the mis-modeling bias for the KS inversion and the lower panel for the SS inversion. We recall that the bias is defined as $(N_\peak^\kappa - N_\peak^g) / N_\peak^g$. The values in the parentheses are estimation errors given by the jackknife technique. The labels \texttt{G1.2}, \texttt{G2.4}, and \texttt{G4.8} respectively stand for Gaussian filters of 1.2, 2.4, and 4.8 arcmin. This table only displays the results from the bin $\nu\in[4.0, 4.5[$.}
    \begin{tabular}{cc|ccc}
        \hline\hline\\[-2.0ex]
        \multirow{2}{*}{$\OmegaM$} & \multirow{2}{*}{$\sigEig$} & \multicolumn{3}{c}{KS inversion} \\
        & & G1.2 & G2.4 & G4.8\\
        \hline\\[-2.0ex]
        0.23 & 0.77 &  12.8\%  (0.9\%) &  27.0\%  (1.4\%) &   3.4\%  (1.8\%) \\
        0.28 & 0.77 &  16.4\%  (0.8\%) &  23.6\%  (1.2\%) &   6.4\%  (1.6\%) \\
        0.33 & 0.77 &  19.0\%  (0.8\%) &  24.0\%  (1.0\%) &   6.7\%  (1.4\%) \\
        0.23 & 0.82 &  15.4\%  (0.8\%) &  22.9\%  (1.2\%) &   5.5\%  (1.6\%) \\
        0.28 & 0.82 &  18.3\%  (0.7\%) &  23.9\%  (1.0\%) &   5.4\%  (1.3\%) \\
        0.33 & 0.82 &  18.2\%  (0.7\%) &  19.7\%  (0.9\%) &   3.1\%  (1.2\%) \\
        0.23 & 0.87 &  16.1\%  (0.8\%) &  19.6\%  (1.0\%) &   1.5\%  (1.3\%) \\
        0.28 & 0.87 &  18.1\%  (0.7\%) &  21.1\%  (0.9\%) &   4.8\%  (1.2\%) \\
        0.33 & 0.87 &  19.0\%  (0.6\%) &  17.7\%  (0.8\%) &   4.7\%  (1.1\%) \\
        \hline\\[-2.0ex]
        \multirow{2}{*}{$\OmegaM$} & \multirow{2}{*}{$\sigEig$} & \multicolumn{3}{c}{SS inversion} \\
        & & G1.2 & G2.4 & G4.8\\
        \hline\\[-2.0ex]
        0.23 & 0.77 &  19.6\%  (1.0\%) &  24.8\%  (1.4\%) &  -0.2\%  (1.7\%) \\
        0.28 & 0.77 &  23.1\%  (0.9\%) &  21.9\%  (1.2\%) &   3.4\%  (1.5\%) \\
        0.33 & 0.77 &  25.7\%  (0.8\%) &  22.7\%  (1.0\%) &   3.6\%  (1.4\%) \\
        0.23 & 0.82 &  22.6\%  (0.9\%) &  21.4\%  (1.2\%) &   2.4\%  (1.5\%) \\
        0.28 & 0.82 &  25.2\%  (0.8\%) &  21.8\%  (1.0\%) &   2.5\%  (1.3\%) \\
        0.33 & 0.82 &  24.5\%  (0.7\%) &  19.0\%  (0.9\%) &   0.9\%  (1.1\%) \\
        0.23 & 0.87 &  22.3\%  (0.8\%) &  18.5\%  (1.0\%) &  -1.9\%  (1.3\%) \\
        0.28 & 0.87 &  24.7\%  (0.7\%) &  20.1\%  (0.9\%) &   2.7\%  (1.2\%) \\
        0.33 & 0.87 &  24.1\%  (0.7\%) &  16.4\%  (0.8\%) &   3.2\%  (1.1\%) \\
        \hline
    \end{tabular}
        \label{tab:bias_filter}
\end{table}

We observe that high and low peaks do not seem to have the same variation. In \fig{fig:peakHist_compInv}, the convergence modeling underestimates the number of low peaks and overestimates high ones. However, their amplitude depends on the applied filter. This is shown more clearly in \tab{tab:bias_filter} where the peak-count bias in a specific bin $\nu\in[4.0, 4.5[$ with all three filter sizes and both inversion methods are presented. The values in the parentheses are the uncertainties of the biases estimated using the jackknife technique. We omitted one of the 2000 realizations at a time and computed the variance over these new derived subsamples. The jackknife errors confirm a bias of the order of 20\% for the two first filters. Neglecting the inversion effect does not affect the bias in the same way for different filters, and we find the same ambiguity for other S/N bins which are not shown.

\begin{figure*}[tb]
        \centering
        \includegraphics[width=8.5cm]{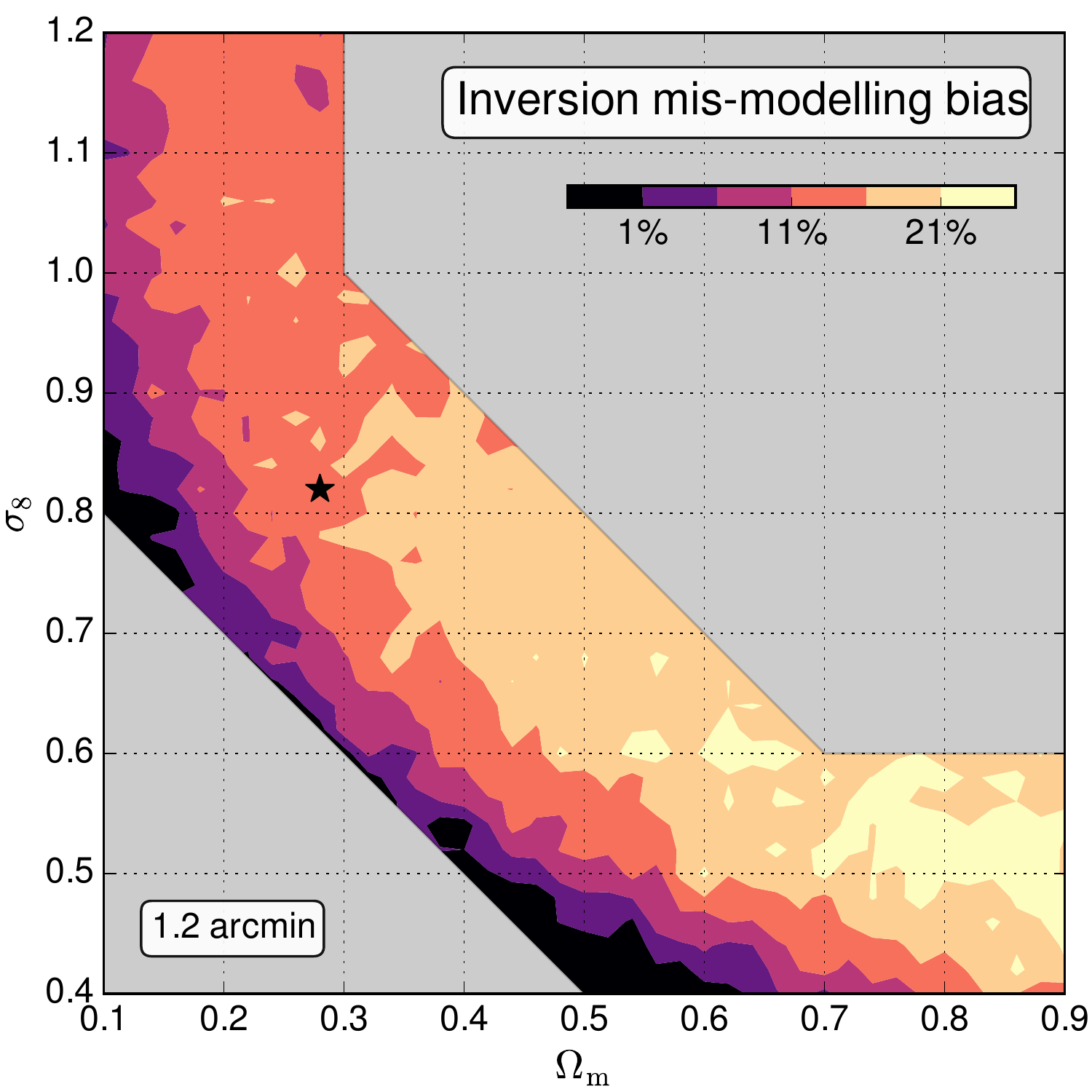}
        \includegraphics[width=8.5cm]{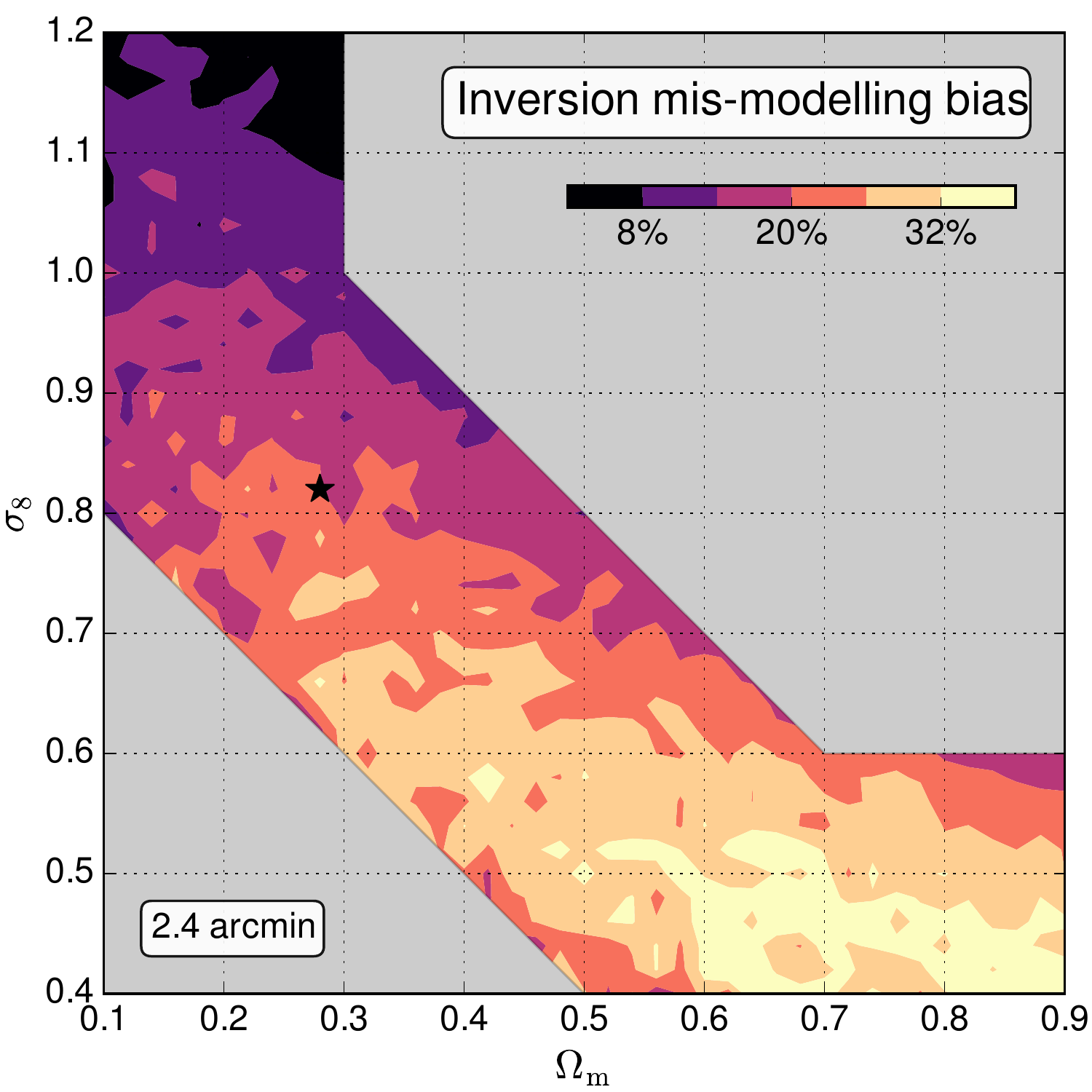}
        \caption{Bias map under a wide range of cosmological parameters. For these two panels, the KS inversion is applied. The dark energy equation of state is $\wZero=-0.96$. The S/N bin is $\nu=[4.0, 4.5[$. The filter size is 1.2 arcmin for the left panel and 2.4 arcmin for the right panel. Contours of parameters sharing the same bias level can be clearly observed. However, they are not necessarily centered on the same parameters when the filter changes. The bias levels are not similar either. These contours vary even more when we examine other choices of bins.}
        \label{fig:bias_case2}
\end{figure*}

\tab{tab:bias_filter} also shows an ambiguous dependence on cosmology. It is not clear that high $\OmegaM$ or high $\sigEig$ would yield a larger or smaller bias. To visualize better the cosmological dependence of the bias, we computed the bias for each cosmology from the second fast simulation set (\sect{sect:methodology}) and obtained a ``bias map'' for the KS inversion (Case 1 compared to Case 2). For example, the bias maps for the bin $\nu\in[4.0, 4.5[$ and the filter sizes of 1.2 and 2.4 arcmin are presented in \fig{fig:bias_case2}. Here, we focus on the $\OmegaM$-$\sigEig$ plane by fixing $\wZero$ at 0.96. Although noisy, the contours of bias levels are visible, and within the studied range of cosmologies, the bias can vary from -4\% to 38\%. The level of noise in \fig{fig:bias_case2} is about two times larger than that of \tab{tab:bias_filter}, since the former possesses four times less realizations than the latter. The figure shows the complex dependence of the peak-count bias on cosmology. Moreover, when we draw the bias maps for other bins and filter sizes, both contour shape and value range vary a lot. This variation is difficult to explain. It seems as if these contours still have a ``banana'' shape but are not centered on the same region at all. Therefore, the reference cosmology can be sometimes in the very biased region, sometimes very far from the very biased region. We argue that cosmology affects the mis-modeling bias in a very complex way that depends on the filter size and the bin range. 

\begin{figure}[tb]
        \centering
        \includegraphics[width=\columnwidth]{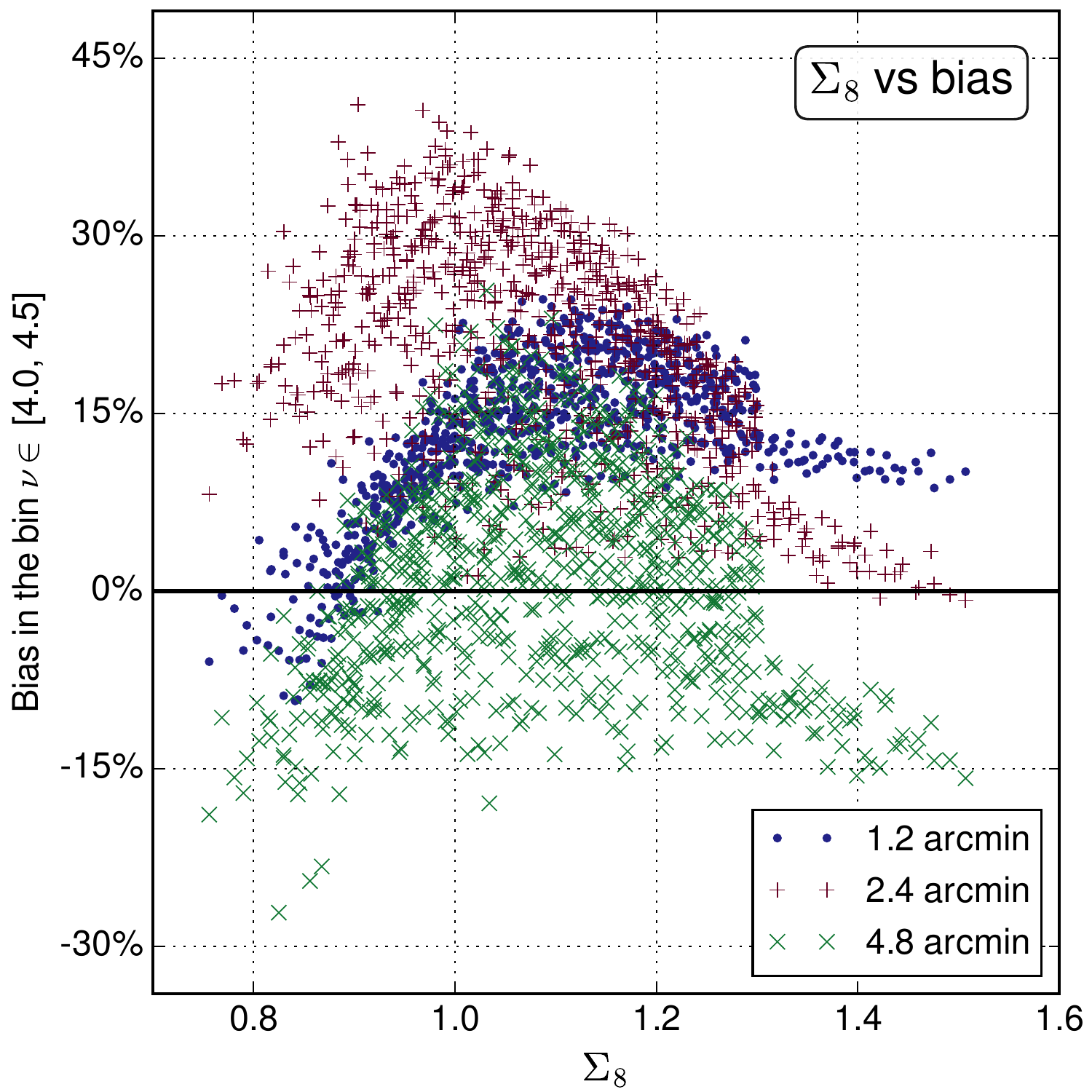}
        \caption{Bias as a function of $\Sigma_8$ for different filters. Blue dots, red pluses, and green crosses are respectively the bias of a specific S/N bin [4.0, 4.5] under different cosmologies for a Guassian filter of 1.2, 2.4, and 4.8 arcmin. Instead of visualizing on the $\OmegaM$-$\sigEig$ plane, here we chose a one-dimensional representation by the reduced parameter $\Sigma_8$. As a result, blues dots contain exactly the same information as the left panel of \fig{fig:bias_case2} and red pluses the same as the right panel.}
        \label{fig:capSigma8_vs_bias}
\end{figure}

Following the geometry of contours displayed in \fig{fig:bias_case2}, we explored the link between the mis-modeling bias and the reduced parameter $\Sigma_8$, defined by \for{for:indicator_1}. \figFull{fig:capSigma8_vs_bias} shows the variation of the bias for different $\Sigma_8$ when three filters are applied. It has the same S/N bin and the same inversion as the previous figure such that the blue dots of \fig{fig:capSigma8_vs_bias} correspond to the left panel of \fig{fig:bias_case2} and red pluses to the right panel. Despite a large scatter, we observe a clear pattern of variations. The filter size and the bin choice should have affected the amplitude of the bias and the value of $\Sigma_8$ where the bias reaches its maximum. The scatter would have been reduced if $\alpha$ or $\beta$ parameters had been chosen differently. However, the optimal values for one case are not necessarily the optimal ones for another. Therefore, we did not look for a minimization of the scatter in \fig{fig:capSigma8_vs_bias}. As a remark, the artifact-like cut at $\Sigma_8=1.3$ for all filters is due to the lack of sample points in the parameter space (gray zone of \fig{fig:bias_case2}). 

\begin{figure}[tb]
        \centering
        \includegraphics[width=\columnwidth]{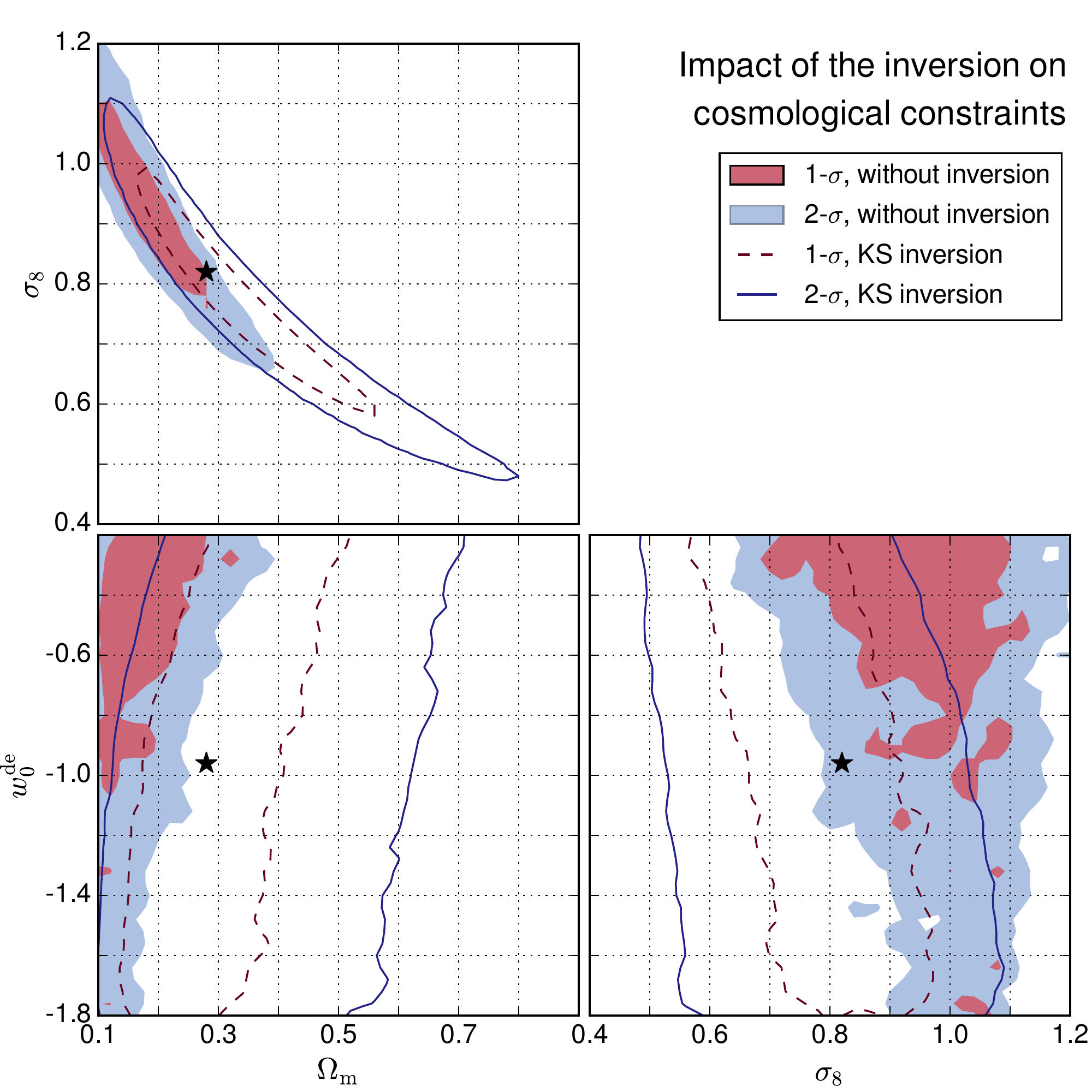}
        \caption{Impact of the inversion mis-modeling bias on $\OmegaM$-$\sigEig$-$\wZero$ constraints. Solid and dashed lines represent contours derived from the inverted shear, whereas shaded areas represent contours derived from the convergence. Although the bias in individual bins is relatively small, the shift of contours is not negligible.}
        \label{fig:contour_corner}
\end{figure}

How does the mis-modeling bias affect cosmological constraints? \figFull{fig:contour_corner} shows a comparison where solid and dashed contours are obtained from the inverted shear of Case 2 and shaded areas from Case 1. More precisely, contours correspond to a case where the inversion has been properly considered, while shaded areas are interpreted as coming from a simplified modeling that ignores the inversion effect. Although the bias in individual bins is relatively small as shown in \fig{fig:peakHist_compInv}, the impact on constraints is not negligible. While the $\OmegaM$-$\sigEig$ constraint shifts approximately along the degeneracy line, both joint constraints with $\wZero$ are more affected, resulting in a biased estimation of parameters. For example, the reference parameter $(\OmegaM, \wZero) = (0.28, -0.96)$ has been excluded at 2$\sigma$ on the left-bottom panel of \fig{fig:contour_corner}.

\begin{table}[tb]
        \centering
        \caption{Diagnostics of cosmological constraints. The indicators $\Sigma_8$, $I_1$, and $I_2$ are respectively defined by Eqs. \eqref{for:indicator_1}, \eqref{for:indicator_2}, and \eqref{for:indicator_3} for each of the two-dimensional projections of the original $\OmegaM$-$\sigEig$-$\wZero$ space.}
        \begin{tabular}{cccc}
                \hline\hline\\[-2.0ex]
                Constraints & $\OmegaM$-$\sigEig$ & $\OmegaM$-$\wZero$ & $\sigEig$-$\wZero$ \\[0.5ex]
                Indicator               & $\Sigma_8$ & $I_1$ & $I_2$ \\
                \hline\\[-1.5ex]
                Case 1 (convergence)    & $1.03^{+0.03}_{-0.04}$ & $0.22^{+0.07}_{-0.04}$ & $0.90^{+0.04}_{-0.15}$ \\[1ex]
                Case 2 (inverted shear) & $1.09^{+0.02}_{-0.03}$ & $0.37^{+0.15}_{-0.07}$ & $0.64^{+0.14}_{-0.11}$ \\[0.5ex]
                \hline
        \end{tabular}
        \label{tab:indicators}
\end{table}

The shift of contours observed above are quantified by the diagnostics mentioned in \sect{sect:methodology}. The values of indicators are shown in \tab{tab:indicators}. On the plane $\OmegaM$-$\sigEig$, \tab{tab:indicators} shows that both cases exclude each other mutually by 1.5--2$\sigma$ , whereas the top panel of \fig{fig:contour_corner} does not suggest such a result. This is due to a bad fitting of $\Sigma_8$ when the dragging parameter $\beta$ is not relaxed. For this family of degeneracy lines, which have a hyperbolic form, there are at least two degrees of freedom. We could have relaxed both $\alpha$ and $\beta$ to make \for{for:indicator_1} loyally describe the degeneracy lines. However, the optimal values of $(\alpha, \beta)$ are not necessarily the optimal ones for other cases, and these parameters have to be set to the same values, otherwise the comparison between $\Sigma_8$ from different cases or surveys would be meaningless. Moreover, there is no physically-motivated choice. Therefore, we were forced to fix $\alpha$ and $\beta$ arbitrarily. Eventually, $\alpha=2/3$ and $\beta=0$ are coherent to most studies in the literature. We also tried the commonly used form $\sigEig(\OmegaM/\text{pivot})^\alpha$ for characterizing $\Sigma_8$ and the same difficulty occurred. This explains why two-dimensional contours between two cases (\fig{fig:contour_corner}) appear more inconsistent with one-dimensional intervals (\tab{tab:indicators}).

From the two lowers panels of \fig{fig:contour_corner}, we find that not considering the mis-modeling bias does yield a 2$\sigma$ tension compared to the realistic case, which justifies our finding that $\wZero$ is more affected. Knowing that the bias depends on the S/N bin and the filter size in a non-trivial way, readers should keep in mind that another choice of the data vector might yield a contour shift different from what we have observed. In the end, the mis-modeling bias is not completely negligible in cosmological constraints.

\section{Summary and discussion}
\label{sect:summary}

In this paper, we performed fast simulations with \Camelus\ to quantify the effect of neglecting the convergence-shear inversion in WL peak-count modeling. We called it ``inversion mis-modeling bias'' since it is introduced when theoretical models fail to include a bias caused by the inversion while it is present in real data. To quantify this effect, on the one hand, we simulated the shear signal as from observations and applied two inversion methods; on the other hand, we simulated directly the convergence signal to count the number of peaks as analytical models do. The comparison of the two yielded an estimation of how theoretical peak models would deviate from the truth.

We have found that not accounting for the inversion has a relatively weak effect on WL peak counts, and that the difference between the KS and SS inversion methods is also small. For the reference cosmology $(\OmegaM, \sigEig, \wZero) = (0.28, 0.82, -0.96)$, the bias from neglecting the inversion is contained within about 30\%. Modeling this bias could be challenging. Its dependency on the S/N value and the filter size is not trivial. Its dependency on cosmology seems to have a link to the reduced parameter $\Sigma_8$, but this link cannot be summarized as a large scatter caused by the choice of the slope $\alpha$ of $\Sigma_8$.

By comparing cosmological constraints obtained from two different cases, we examined the propagation of the mis-modeling bias during the process of estimating cosmological parameters. We have found that the $\OmegaM$-$\sigEig$ contours are less affected. The change is mainly along the degeneracy line. However, for the dark energy equation of state $\wZero$, we have seen that both on the $\OmegaM$-$\wZero$ and $\sigEig$-$\wZero$ planes, contours can exclude each other by 2$\sigma$. The dark energy parameter seems to be more affected.

For the future WL surveys aimed at constraining cosmological parameters with high precision, modeling any possible systematic sources, including the inversion, will be indispensable. For cosmic shear, the inversion can be bypassed since linking the two-point-correlation functions of the shear to theory is straightforward. For peaks, unless we only focus on the very high S/N regime, the link is less trivial. A possible option to bypass the inversion would be the aperture mass \citep{Kaiser_etal_1994, Schneider_1996}, which allows us to identify tangential shear peaks. However, we should remember that the aperture mass does not account for the nonlinearity of the reduced shear. Either using the aperture mass or the inversion, modeling the correction should be a common goal for future studies.

\begin{acknowledgements}
        For their financial support, the authors would like to thank the DIM-ACAV thesis fellowship of the Région d'Île-de-France and the French national program for cosmology and galaxies (PNCG), and the computing support from the in2p3 Computing Centre. We are also thankful to the anonymous referee for highlighting ambiguities of writing and for her/his time investment. We also thank Zuhui Fan and Xiangkun Liu for discussions about this paper. Linc is grateful to François Lanusse and Sandrine Pires for discussions on the build-up of tests used in this paper, to Marie Gay for her endless help on technical issues, and to Li-Wen Huang, Corentin Champseix, Pin-Ting Lin, and Yu-Jung Sun for their aid in Linc's relocation to Edinburgh.
\end{acknowledgements}

\bibliography{%
BibList_WL,%
BibList_WLPC,%
BibList_TR%
}

\end{document}